\documentclass[twoside,slac_one]{revtex4}
\usepackage{graphicx}
\usepackage{fancyhdr}
\usepackage{amsmath} 
\usepackage{bm}
\usepackage{amsxtra}
\usepackage{amssymb}
\usepackage{amsthm}
\usepackage{latexsym}
\usepackage{lscape}

\begin{document}

\title{Data Quality and Performance of the NO\boldmath{$\nu$}A Prototype Detector}

%

\author{S. Lein for the NO$\nu$A collaboration}
\affiliation{Department of Physics and Astronomy, University of Minnesota, Minneapolis, MN, USA}

\begin{abstract}
The NuMI Off-Axis $\nu_{e}$ Appearance (NO$\nu$A) project is a long-baseline neutrino experiment. It utilizes the NuMI
neutrino beam at Fermilab and consists of two functionally-identical liquid scintillator filled detectors. The detectors are placed 14
milliradians off-axis from the beam and 810 km apart. A 209 ton prototype detector, the Near Detector On the
Surface (NDOS), was built and began taking initial neutrino data in December 2010. NDOS is 110 milliradians off-axis from
the NuMI beam and also records neutrinos from the Booster Neutrino Beam. As NDOS is in the commissioning phase,
metrics are being developed to improve understanding of the detector as well as monitor the quality of data. Performance
of the prototype detector will be presented.

\end{abstract}

\maketitle

\thispagestyle{fancy}


\section{The NO\boldmath{$\nu$}A Experiment}

\begin{figure}[h]
\centering
\includegraphics[width=80mm]{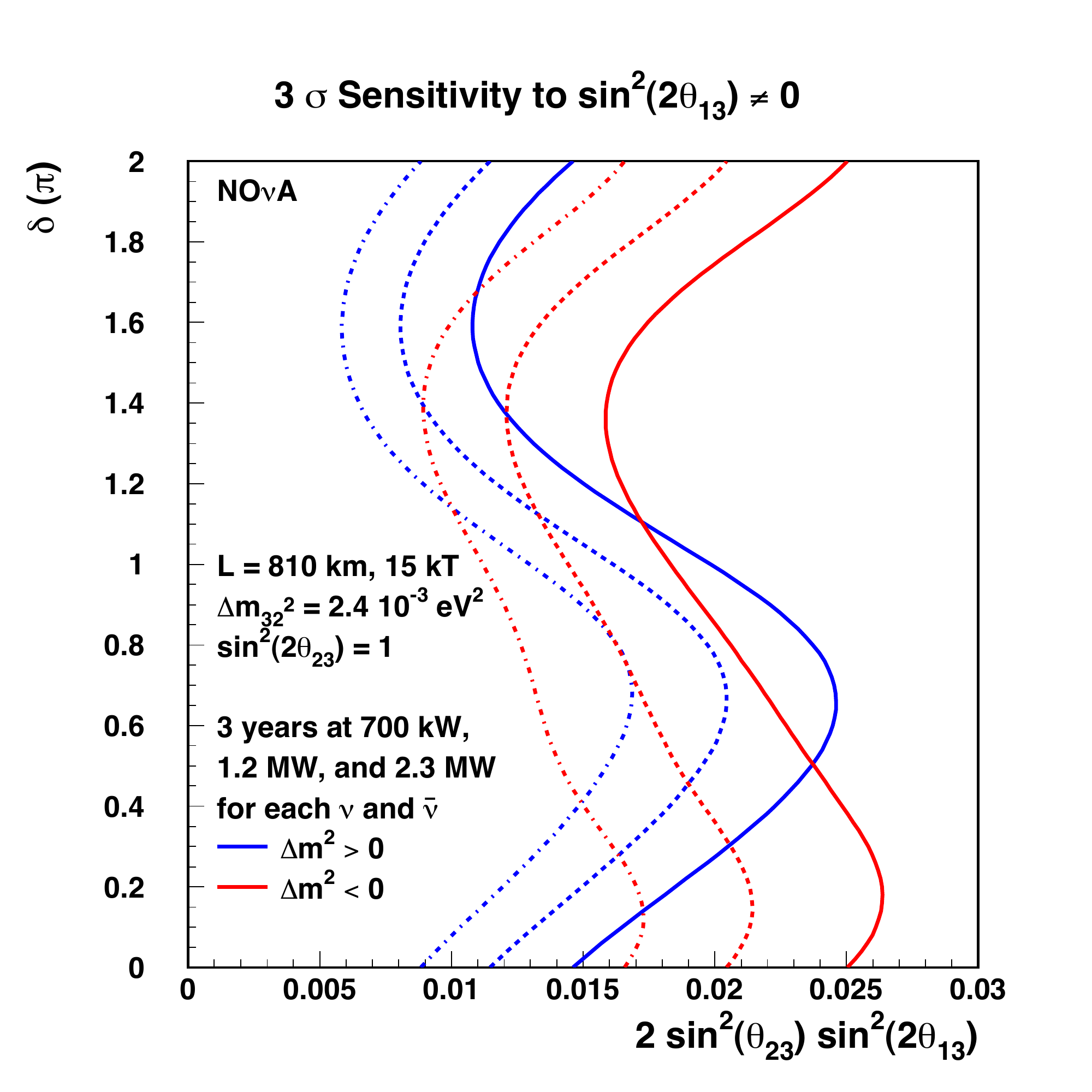}
\caption{NO$\nu$A 3$\sigma$ sensitivity as a function of $\theta_{13}$. The blue curves corresponds to normal mass hierarchy, while the red curves correspond to the inverted mass hierarchy. These sensitivities are based on 3 years of running each with a neutrino and anti-neutrino beam. The dotted curves represent higher beam intensities but the solid curves are with the planned 700 kW beam intensity. } \label{theta13}
\end{figure}

The NuMI\footnote{Neutrinos at the Main Injector\cite{numi}} Off-axis $\nu_{e}$ Appearance (NO$\nu$A) experiment will be able to address many questions about the parameters of neutrino oscillation. NO$\nu$A will be an order of magnitude more sensitive to values of $\sin^{2}(2\theta_{13})$ than current limits near 0.1, as shown in Figure \ref{theta13}.  It will be able to probe the mass hierarchy as well as examine if oscillations violate Charge-Parity. NO$\nu$A will also make precision measurements of $\theta_{23}$ and $|\Delta m_{32}^2|$.

NO$\nu$A consists of two detectors, one located at Fermilab in Batavia, IL, (called the near detector) and one located at Ash River, MN near the Canadian border and Voyageurs National Park (called the far detector). These detectors are functionally-identical except for overall size. The near detector will be about 200 tons, while the far detector will be about 14 kton. These detectors are placed to intercept the neutrinos from the NuMI beam created at Fermilab\cite{TDR}.

The NO$\nu$A detectors have a cellular structure. Each cell is a reflective PVC tube that is filled with liquid scintillator (mineral oil doped with about 5$\%$ pseudocumene). Within the cell is a loop of wave-length shifting fiber whose ends are connected to an Avalanche Photodiode (APD). For the far detector, the dimensions of the cell are 4 cm x 6 cm x 15 m. The near detector cells have the same cross-sectional area, but are 4 m long. These dimensions result in 0.15 radiation lengths per plane of cells. The planes of cells are orientated perpendicular to the direction the neutrinos are traveling and each plane alternates from being horizontally or vertically aligned in the perpendicular plane. This allows for full three-dimensional reconstruction.

When a neutrino interacts within the detector volume, charged particles are created. As charged particles travel through the liquid scintillator, they create photons. These photons are absorbed by the wave-length shifting fiber and are transported to the fiber end, where they are read-out by the APD. The APD amplifies the signal by a factor of one hundred, a level capable of being read by sensitive, low-noise electronics. 

Each APD has 32 pixels which allows it to read out one module. Each APD is then connected to a Front End Board (FEB). The FEB digitizes the signal from the APD and sends it to a Data Concentrator Module (DCM). Each DCM is capable of reading out up to 64 FEBs. This signal is then passed to the rest of the data acquisition system\cite{TDR}. 

\section{The Near Detector On the Surface (NDOS)}
The NDOS is the prototype detector for the NO$\nu$A detectors. At 209 tons, it is approximately the size of the future near detector and is functionally-identical. It is located on the surface at Fermilab instead of underground, as the near detector will be. The NDOS is currently taking neutrino data. Having a prototype of this scale has proved invaluable in preparation for construction of the real detectors. It has allowed us to test and improve everything from assembly procedures to the timing systems. 

NDOS detects neutrinos from the NuMI beam, currently running for the MINOS experiment\cite{minos}, as well as from the Booster Neutrino beam. The Booster Neutrino beam is primarily for the MiniBooNE experiment\cite{miniboone}. The NDOS detector is 110 mrad off-axis from the NuMI beam and the detector axis is in the same plane as the beam axis. The NDOS is nearly on-axis relative to the Booster beam and the detector axis is rotated relative to the beam axis by $23^{\circ}$. 


\section{Monitoring Systems}
The NDOS has five levels of data monitoring: the Data Acquisition (DAQ) Monitor, the Memory Viewer, the Event Display, the Online Monitor, and DataCheck. The first four are real-time monitors, while DataCheck has a delay of about one hour. Both Online Monitor and DataCheck report run metrics. Under stable running conditions, runs are about a day in length, with subruns lasting approximately one hour. Subruns consist of trigger windows. Everytime the system triggers on either a neutrino spill or a regular, 10 Hz cosmic trigger, a 500 $\mu$sec time window of data is written out. The near and far detectors will have much shorter trigger windows; the longer windows for NDOS allowed us to increase the load on the DAQ system.

\subsection{DAQ Monitor}
\begin{figure*}[t]
\centering
\includegraphics[width=135mm]{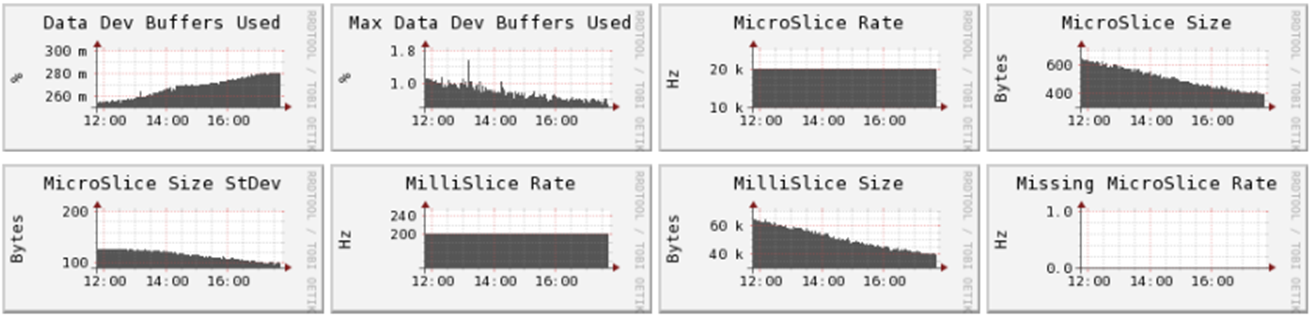}
\caption{An example of plots generated by the DAQ Monitor to monitor a DCM. For all the plots, the horizontal axis is current time in hours. \textit{Data Dev Buffers Used} and \textit{Max Data Dev Buffers Used} plot the percentage of circular buffer capacity for the DCM device driver that was occupied on average and at the peak for each time interval. Monitoring these values ensures we are not using all the available capacity, which could result in lost data. \textit{MicroSlice Rate} is the rate in Hz of microslices for each time interval. Microslices are concatenated amounts of data and the rate should be steady, since it is a configured parameter. The \textit{MicroSlice Size} and \textit{MicroSlice StDev} report the average size in bytes and standard deviation of the size of microslices. The size is a reflection on the amount of data we are collecting; the standard deviation is a measure of how steady this data rate is. Millislices are larger concatenations of data. We monitor the \textit{MilliSlice Rate} and \textit{MilliSlice Size} in a manner analogous to microslices. Lastly, we monitor the \textit{Missing MicroSlice Rate} in Hz. This should be zero if we are not losing data. } \label{ganglia}
\end{figure*}

The first level of monitoring is done by the DAQ Monitor. It monitors the health and performance of the entire DAQ System, from the DCMs to the network connections. It uses Ganglia as its base. Ganglia is a third-party, easily customizable, open-source software that specializes in tracking statistics like memory and network usage. NO$\nu$A has created metrics such as data and trigger rates, data sizes, error states, and the level of data corruption. All of these metrics are written to a database and accessed through a web interface which allows them to be plotted interactively. The example display shown in Figure \ref{ganglia} is used to monitor the status and performance of a DCM.  In general, custom monitored metrics can be used to troubleshoot problems, optimize performance, and diagnose trends over both short and long time periods.

\subsection{Memory Viewer}

Memory Viewer is the next level of monitoring. Memory Viewer displays colored bytes of the raw data from the detector, giving an intuitive sense of the data flow. Using pattern matching, one can tell if the run is operating normally. Memory Viewer, as well as many of the other monitoring systems, relies on Event Dispatcher. The Event Dispatcher is a server program running on a DAQ host which streams event data to clients, which connect to it via a TCP socket. Memory Viewer connects to this stream and displays the bytes.

\begin{figure*}[t]
\centering
\includegraphics[width=135mm]{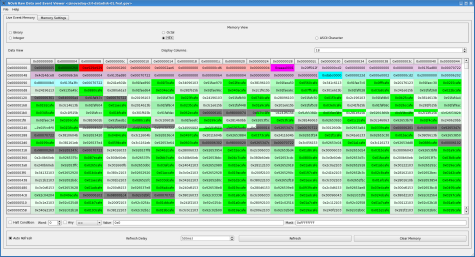}
\caption{This is the graphical user interface of the Memory Viewer. The raw data is displayed as colored bytes, giving an intuitive sense of the data. Using pattern matching, one can quickly assess the run status.} \label{memoryviewer}
\end{figure*}

\subsection{Event Display}

\begin{figure*}[t]
\centering
\includegraphics[width=135mm]{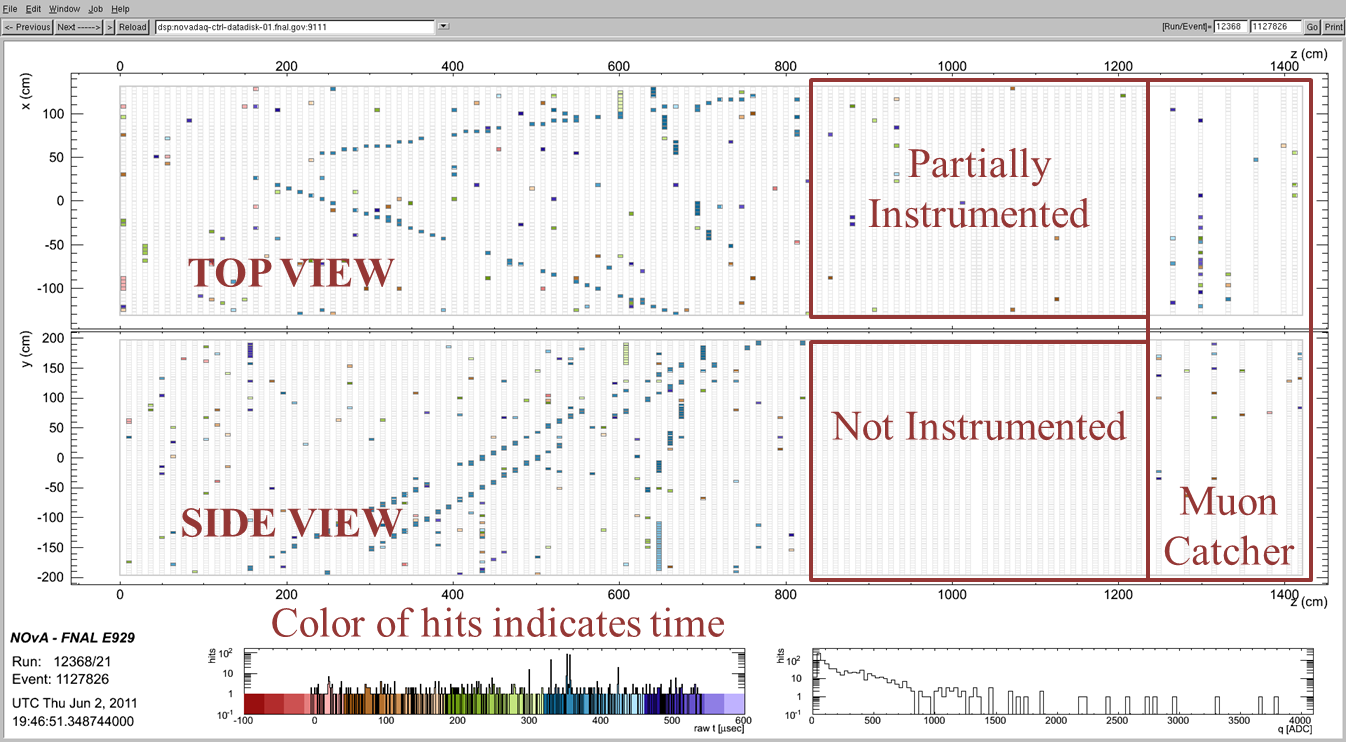}
\caption{The Event Display is a detector view of the data. It shows both the top and side views of the detector. Each channel is a grey rectangle. If the channel had a hit, it is colored. In this case, the color indicates the relative time of the hits. The beam enters the front of the detector on the left-hand side of the Event Display. The detector has regions that are uninstrumented or only partially instrumented. The muon catcher is located at the back of the detector and has layers interspersed with steel planes.} \label{evd}
\end{figure*}

Event Display is a detector view of the trigger windows from Event Dispatcher. It shows both of the two-dimensional views: the view from the top of the detector and the view from the side. Each channel is represented in one of the views as a grey rectangle. If the channel had a hit during the trigger window, it is colored. The color can indicate the relative time of the hit, as in Figure \ref{evd}, or the relative charge of the hit. The beam enters the front of the detector, the left-hand side of the Event Display. Towards the back of the detector, we have regions that are either uninstrumented or only partially instrumented. Finally, the muon catcher is represented at the right side of the Event Display; its planes are interspersed with steel so the instrumentation is less dense. Figure \ref{evd} shows cosmics rays entering the detector.

\subsection{Online Monitor}

\begin{figure*}[t]
\centering
\includegraphics[width=135mm]{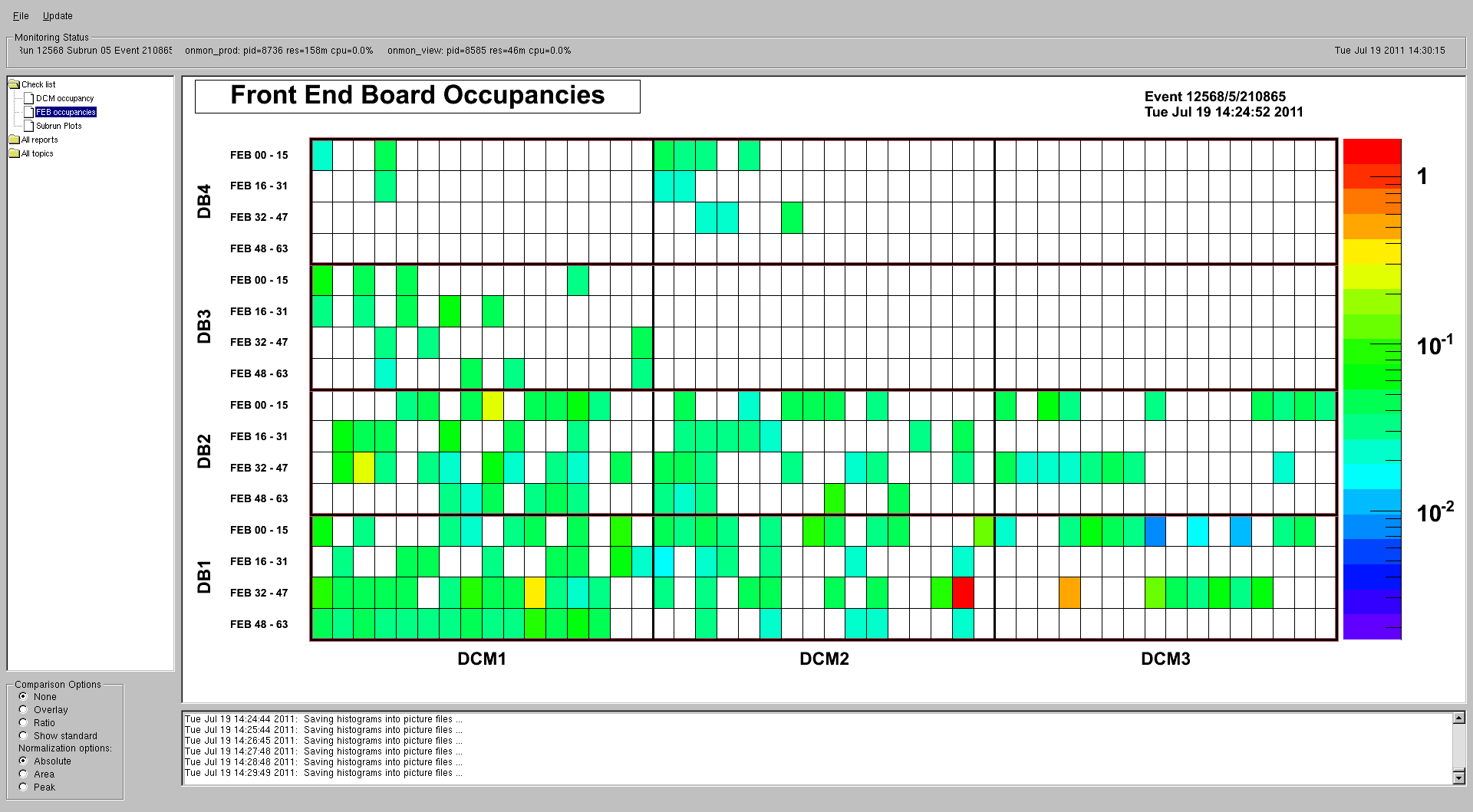}
\caption{Plot of FEB occupancies from Online Monitor. Each rectangle represents one FEB and its color is the relative numbers of hits as given by the scale at the right of the figure. This gives an indication of each FEB's noise levels.} \label{onmon}
\end{figure*}

Online Monitor displays run metrics such as the number of active channels and channel occupancies in real time. Figure \ref{onmon} shows Online Monitor displaying the occupancies of the FEBs which gives an indication of which FEBs are noisier than others. Online Monitor is broken into two processes to increase system stability. Online Monitor Producer reads in information from the Event Dispatcher and processes it to create run metrics. Online Monitor Viewer is a graphical user interface that communicates with the Producer through a ROOT TMapFile. The Viewer is responsible for actually displaying the information. 

\subsection{DataCheck}

DataCheck is the final level of monitoring. It is an offline software tool and has a delay of about one hour from the end of a subrun. DataCheck has the ability to look at metrics subrun by subrun, or cumulatively across multiple subruns or runs to find trends in the data.  It also has a web interface that uses PHP to make plots from a database. DataCheck monitors metrics, such as the number of active channels, the average number of hits per channel, the number of active FEBs, and the relative times of neutrino candidates. Section \ref{performance} reports some of the findings from DataCheck.

\section{Performance of the NDOS\label{performance}}

\begin{figure*}[t]
\centering
\includegraphics[width=135mm]{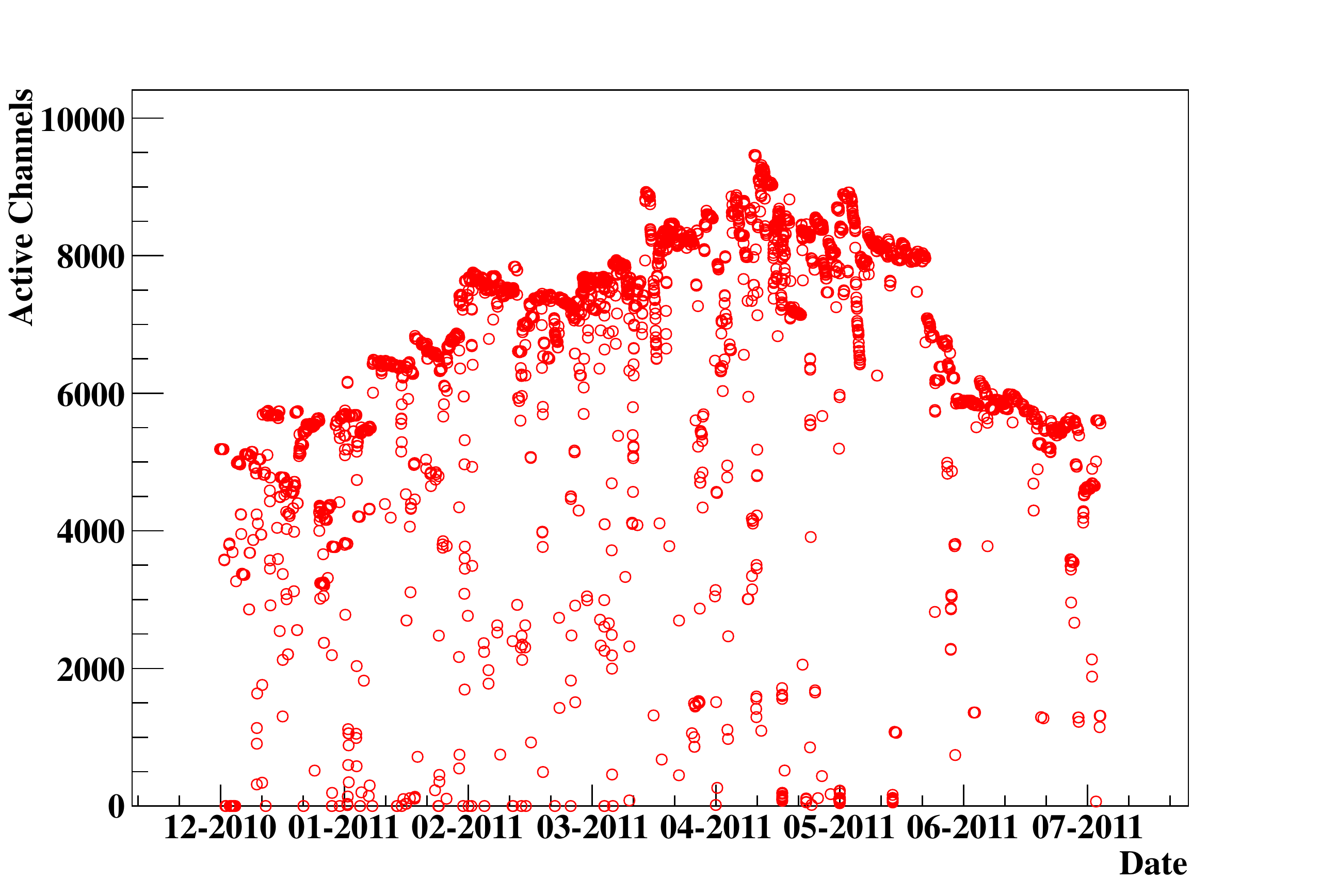}
\caption{Number of active channels in NDOS. In May 2011, we began actively removing questionable channels for further study. Intermittent drops are due to running with a partial detector during commissioning. A fully-instrumented NDOS would have 15,900 channels.} \label{chan}
\end{figure*}

One fundamental metric of the performance of the NDOS is the number of active channels. Active channels are defined as those that are instrumented and operating within acceptable parameters. From Figure \ref{chan}, one can see that channels were steadily installed throughout the winter of 2010 and spring of 2011. However, in May 2011, we began actively removing all channels that displayed questionable behavior. This has allowed us to fully understand all the problems with our channels before we begin procurement for the far detector. One may also note the drops in channel level throughout the run period. This is due to frequent running with only parts of the detector while work is being done on other sections. For reference, a fully-instrumented NDOS would have 15,900 channels.

\begin{figure*}[t]
\centering
\includegraphics[width=135mm]{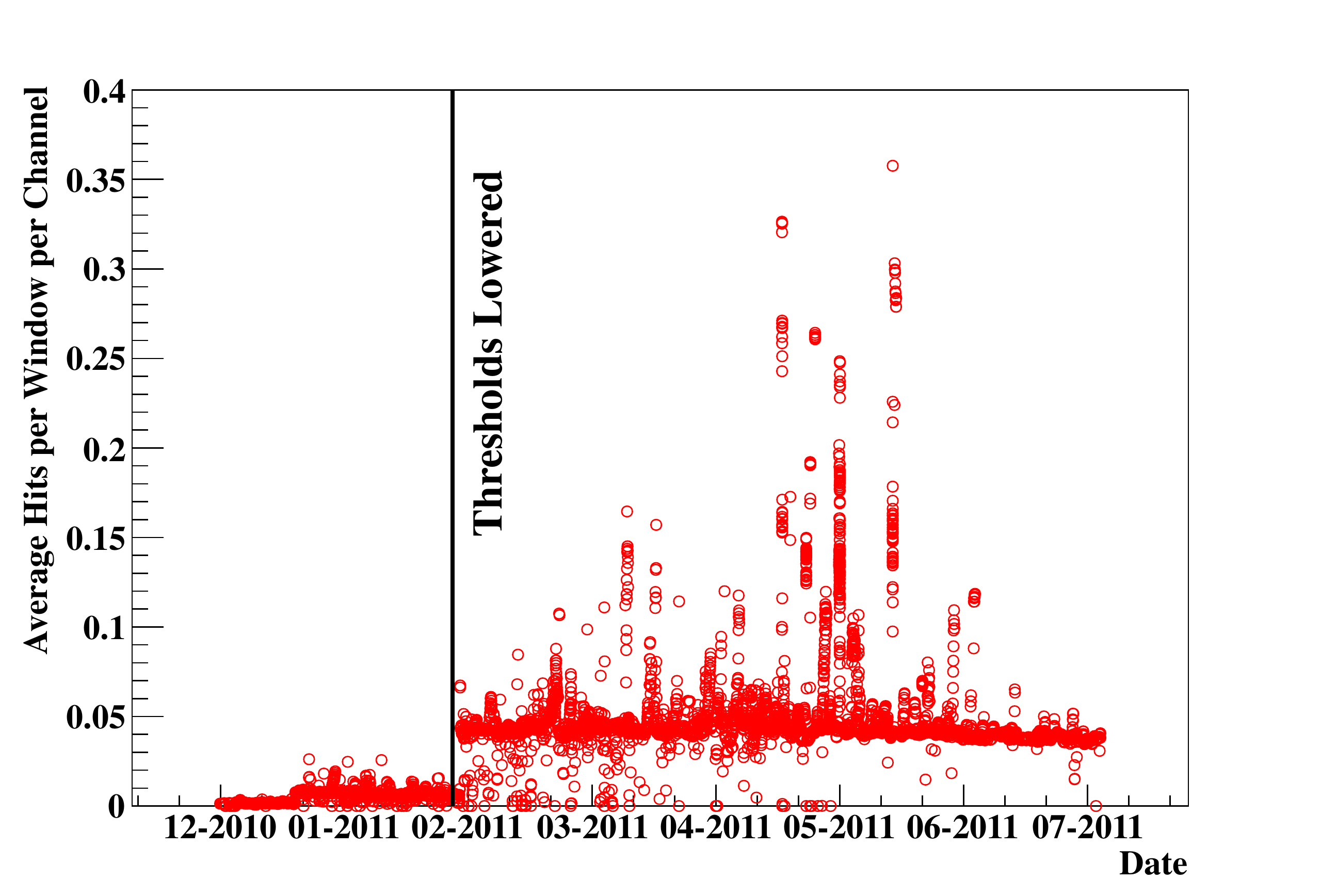}
\caption{Hits per trigger window (500 $\mu$sec) per active channel. Initial running had high thresholds and thus low values of this metric. After lowering the thresholds, we have been relatively stable at 5$\%$. Jumps are due to noisy periods in the NDOS that are under investigation. } \label{hits}
\end{figure*}

Another basic metric is the number of hits per trigger window per active channel. One can see from Figure \ref{hits} that it is generally 5$\%$. When initally commissioning NDOS, we set high thresholds and the average number of hits was lower. Once we understood our system, we lowered our thresholds to more reasonable levels, which accounts for the rise in average number of hits. One can see that sometimes the NDOS gets very noisy. This is due to humidity effects, electronics noise, etc. These effects are under study and we are working to mitigate them in the near and far detectors.

This noisy behavior is also reflected in the rate of FEB shutoffs. Our system is designed to shutoff a FEB during a run if the data buffer overflows. The FEB then remains off for the rest of the run and is re-initialized when the next run begins. Data buffers usually overflow due to noisy APDs. Thus, high rates of FEB shutoffs are a concern. Initially, we noticed noisy run conditions seemed to be correlated with high humidity in the detector hall. Then we began to cool our APDs to $-15^{\circ}$C to reduce the noise levels. However, when we did this on the NDOS, some channels became unusable. It was found that many had water on the surface of the APD due to condensation from air infiltration. We are investigating and redesigning our electronics installation procedure and circuit boards to ensure that the APDs are in a sealed environment.

\begin{figure*}[t]
\centering
\includegraphics[width=135mm]{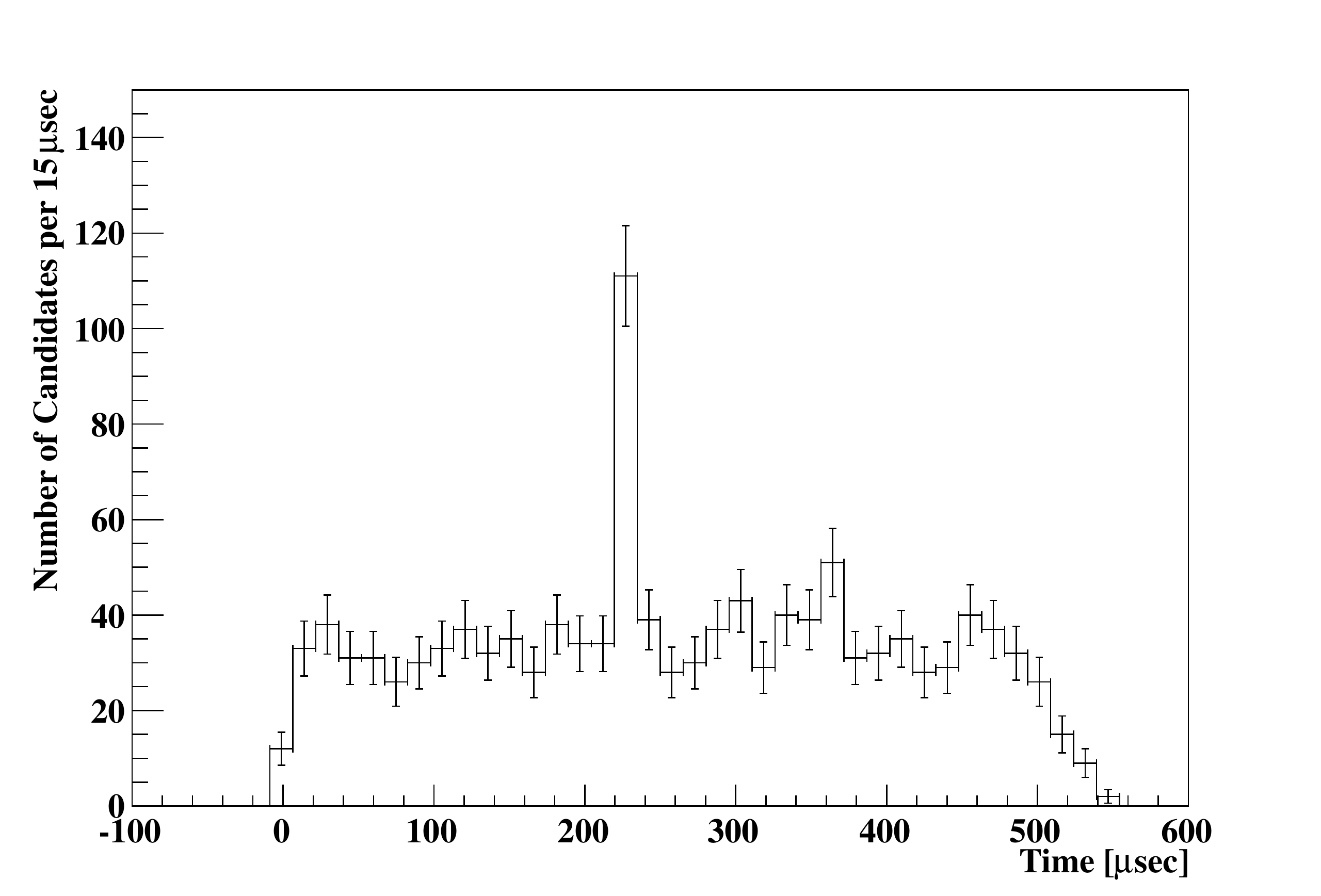}
\caption{Times of NuMI neutrino candidates within the trigger window (500 $\mu$sec) for one week's running. Neutrino candidates are found using rough activity, fiducial, and direction cuts. If a peak is seen in the distribution within the 10 $\mu$sec NuMI beam spill, we have verified that the trigger and timing systems are working as expected.} \label{numi}
\end{figure*}

Using rough activity, fiducial, and direction cuts, DataCheck can quickly find neutrino candidates. The time distribution of these candidates relative to the trigger window is plotted in Figure \ref{numi}. A peak in this distribution within the 10 $\mu$sec NuMI beam spill indicates the timing and triggering systems are operating as expected. Figure \ref{numi} contains one week's worth of running and a peak is clearly visible. This allows us to verify if our systems are working on a reasonably short time scale. We also track the Booster neutrino peak; however, with the lower statistics, it takes more than a week to see a peak.

\section{Summary}

In summary, the NDOS has been taking data for 9 months. The NO$\nu$A collaboration will begin building the near and far detectors this winter. We have viable systems in place to monitor both detector and data quality. These systems continue to aid us in commissioning NDOS and preparing for future detector construction.

\bigskip 

\end{document}